\documentstyle[aps]{revtex}

\title{Magnetic Catalysis of Dynamical Symmetry Breaking and
Aharonov-Bohm Effect}

\author{V. A. Miransky\cite{email}}

\address{Bogolyubov Institute for Theoretical Physics,\\
252143 Kiev, Ukraine\\
and\\
Department of Physics\\
Nagoya University\\
Nagoya 464-01, Japan}

\draft
\begin{document}
\maketitle

\bibliographystyle{unsrt}

\arraycolsep1.5pt

\begin{abstract}
The phenomenon of the magnetic
catalysis of dynamical symmetry breaking is based on
the dimensional reduction $D\to D-2$ in the dynamics of
fermion pairing in a magnetic field. We discuss 
similarities between this phenomenon and the Aharonov-Bohm
effect. This leads to the interpretation of the
dynamics of the (1+1)-dimensional Gross-Neveu model with
a non-integer number of fermion colors as
a quantum field theoretical analogue of the
Aharonov-Bohm dynamics.
\end{abstract}

\section{Introduction.}

It has been recently shown \cite{1,2,3,4} that a constant magnetic
field in $2+1$ and $3+1$ dimensions is a strong catalyst of dynamical
chiral symmetry breaking, leading to the generation of a fermion
dynamical mass even at the weakest attractive interaction between
fermions (the magnetic catalysis of dynamical 
symmetry breaking).

The effect has been extensively studied
in different models \cite{ng,el,ngT,gs}, 
confirming the universality of this
phenomenon.\footnote{The
fact that an external magnetic field enhances a fermion
dynamical mass was known from studying the NJL model in 3+1 and
2+1 dimensions \cite{klev,krive}.}

There might be interesting applications of this effect in
condensed matter physics \cite{nick,ssw} and 
cosmology \cite{3,4}.\footnote{As it has been shown in Refs.
\cite{ngT,gs}, at weak coupling 
constants of quantum dynamics (such as gauge and Yukawa interactions)
the magnetic catalysis is irrelevant for the phase transitions in the
early Universe.
However, it may become relevant if some coupling constants are
strong, as technicolor interactions during the electroweak phase
transition (compare with Ref.\cite{inc1}).}
As it has been recently pointed out
\cite{itoh}, the effect should play the important role in the
phenomenon of the spontaneous generation of a magnetic field in
the vacuum of 2+1 dimensional QED with the Chern-Simons
term \cite{hosot}. 

The essence of the effect is the dimensional reduction $D\to 
D-2$ (i.e. $2+1 \to 0+1$ and $3+1 \to 1+1$ ) in the infrared 
dynamics of the fermion pairing in a magnetic field
\cite{1,2,3,4}.
The physical reason of this reduction is
the fact that the motion of 
charged particles is restricted in those directions that are 
perpendicular to the magnetic field.  This is in turn connected 
with the point that, at weak coupling between fermions, the 
fermion pairing, leading to the chiral condensate, is mostly 
provided by fermions from the lowest Landau level (LLL) whose 
dynamics are ($D-2$)--dimensional. 

The most explicit description of this dimensional reduction was
done by Elias et al.\cite{el}.
In that  paper, it was shown that in the ``continuum" 
limit, when both the strength of the magnetic field and the 
ultraviolet cutoff go to infinity, the (weakly coupling)
$(3+1)$-dimensional NJL models with $N_c$ 
colors are reduced to a continuum set of independent 
(1+1)--dimensional Gross-Neveu (GN) models \cite{5}, labeled by 
coordinates ${\bf x}_{\perp}$ in the plane perpendicular to the 
magnetic field ${\bf B}$.  The number of colors in the GN models 
is $\tilde{N}_c=( b\pi/2)N_c$, where $b$ is $b=|eB|/\Lambda^2$ in 
the ``continuum" limit (here $\Lambda$ is the ultraviolet 
cutoff).  The factor $b\pi/2$ is 
proportional to a (local) magnetic flux attached to each point 
in the $x_{\perp}$-plane (see Sec.3 below). Actually, $b\pi/2$ is
equal to $|e\Phi|/2\pi$, where $\Phi$ is the local magnetic flux,
and therefore
\begin{equation}
\tilde{N}_c=N_c (|e\Phi|/2\pi).
\label{eqN_c}
\end{equation}
If we asserted that $\tilde{N}_c$ is a (non-negative) integer,
we would be led to the
quantization condition for $N_c(e\Phi/2\pi)$ coinciding with
the quantization condition for a magnetic flux,
$q(\Phi/2\pi)=n$, 
at which the Aharonov-Bohm scattering 
of particles with the electric charge $q=eN_c$
disappears \cite{AB}.

We will argue that this intriguing similarity reflects a deep
connection between the phenomenon of
the dimensional reduction in a magnetic field
and the Aharonov-Bohm effect. This in turn yields the
interpretation of the dynamics in the
GN model with a non-integer number
of colors $\tilde{N}_c$ as a quantum field theoretical
analogue of the Aharonov-Bohm dynamics
with a non-integer $q(\Phi/2\pi)$.

But first, following Ref.\cite{el},
we shall discuss the connection between the (1+1)-dimensional
GN model and the (3+1)-dimensional NJL model in a magnetic
field.

\section{Effective Action in the Gross--Neveu Model.}

In this section, for completeness, we shall derive the effective 
action for the GN model. The Lagrangian density of the GN model 
is:
\begin{equation}
{\cal L}_{GN} = \frac{1}{2} 
\left[\bar{\Psi}, (i\gamma^\mu \partial_\mu)\Psi\right] +
\frac{\tilde{G}}{2} 
\left[ (\bar{\Psi}\Psi)^2+(\bar{\Psi}i\gamma^5\Psi)^2 \right]
\label{eq1}
\end{equation}
where $\mu=0,1$ and the fermion field carries an additional 
``color" index $\tilde{\alpha}=1,2,\dots,\tilde{N}_c$ (for 
simplicity, we consider the case of the chiral $U_L(1)\times 
U_R(1)$ symmetry). The theory is equivalent to the theory with 
the Lagrangian density
\begin{equation}
{\cal L}^{\prime}_{GN}=\frac{1}{2} 
\left[\bar{\Psi}, (i\gamma^\mu  \partial_\mu) \Psi\right] 
- \bar{\Psi}(\sigma+i\gamma^5\pi)\Psi
- \frac{1}{2\tilde{G}} \left(\sigma^2+\pi^2\right).
\label{eq2}
\end{equation}
The Euler--Lagrange equations for the auxiliary fields $\sigma$ and
$\pi$ take the form of constraints:
\begin{equation} 
\sigma=-\tilde{G}\bar{\Psi}\Psi, \qquad 
\pi=-\tilde{G}\bar{\Psi}i\gamma^5\Psi,
\label{eq3}
\end{equation}
and the Lagrangian density (\ref{eq2}) reproduces Eq.(\ref{eq1}) upon
application of the constraints (\ref{eq3}). The effective action for
the composite fields $\sigma$ and $\pi$ can be obtained by
integrating over fermions in the path integral. It is given by the
standard relation: 
\begin{eqnarray}
\Gamma_{GN}(\sigma,\pi) &=& \tilde{\Gamma}_{GN}(\sigma,\pi)
-\frac{1}{2\tilde{G}}\int d^2x(\sigma^2+\pi^2), 
\label{eq4} \\
\tilde{\Gamma}_{GN}(\sigma,\pi) &=&- i Tr Ln \left[i\gamma^\mu
\partial_\mu - (\sigma+i\gamma^5\pi)\right].
\label{eq5}
\end{eqnarray}
The low energy quantum dynamics are described by the path 
integral (with the integrand $\exp(i\Gamma_{GN})$) over the 
composite fields $\sigma$ and $\pi$. As $\tilde{N}_c\to\infty$, 
the path integral is dominated by the stationary points of the 
action:  
$\delta\Gamma_{GN}/\delta\sigma=\delta\Gamma_{GN}/\delta\pi=0$. 
We will analyze the dynamics by using the expansion of the 
action $\Gamma_{GN}$ in powers of derivatives of the composite 
fields.

We begin the calculation of $\Gamma_{GN}$ by calculating the 
effective potential $V_{GN}$. Since $V_{GN}$ depends
only on the $U_L(1)\times U_R(1)$--invariant $\rho^2=\sigma^2+\pi^2$, 
it is sufficient to consider a configuration with $\pi=0$ and $\sigma$ 
independent of $x$. Then we find from Eqs.~(\ref{eq4}) and (\ref{eq5}):
\begin{eqnarray}
V_{GN}(\rho)&=&\frac{\rho^2}{2\tilde{G}} 
             - \tilde{N}_c\int \frac{d^2k}{(2\pi)^2} 
               \ln\left(\frac{k^2+\rho^2}{k^2}\right)=
\nonumber\\
            &=&\frac{\rho^2}{2\tilde{G}} 
             - \frac{\tilde{N}_c\rho^2}{4\pi}
               \left[\ln\frac{\Lambda^2}{\rho^2}+1\right],
\label{eq6}
\end{eqnarray}
where the integration is done in Euclidean region ($\Lambda$ is an
ultraviolet cutoff). As is known, in the GN model the equation of
motion $dV_{GN}/d\rho=0$ has a nontrivial solution
$\rho =\bar{\sigma}\equiv m_{dyn}$ for any value of the coupling
constant $\tilde{G}$. Then the potential $V_{GN}$ can be rewritten as 
\begin{equation}
V_{GN}(\rho)=\frac{\tilde{N}_c\rho^2}{4\pi}
\left[\ln\frac{\rho^2}{m_{dyn}^2}-1\right],
\label{eq7}
\end{equation}
where
\begin{equation}
m_{dyn}^2=\Lambda^2\exp
\left(-\frac{2\pi}{\tilde{N}_c\tilde{G}}\right).
\label{eq8}
\end{equation}
Due to the Mermin-Wagner-Coleman
(MWC) theorem \cite{7}, there cannot be spontaneous
breakdown of continuous symmetries at $D=1+1$. The parameter $m_{dyn}$
is an order parameter of chiral symmetry breaking only in leading
order in $1/\tilde{N}_c$ (this reflects the point that the MWC
theorem is not applicable to systems with $\tilde{N}_c\to \infty$
\cite{8}). In the exact GN solution, spontaneous chiral symmetry
breaking is washed out by interactions (strong fluctuations) of
would--be NG bosons $\pi$ (i.e. after integration over $\pi$ and 
$\sigma$ in the path integral). The exact solution in 
this model corresponds to the realization of the
Berezinsky--Kosterlitz--Thouless (BKT) phase: though chiral
symmetry is not broken in this phase, 
the parameter $m_{dyn}$ still defines the
fermion mass, and the would--be NG boson $\pi$ transforms into a BKT
gapless excitation \cite{8}.

Let us now turn to calculating the kinetic term in $\Gamma_{GN}$. The
chiral $U_L(1)\times U_R(1)$ symmetry implies that the general form
of the kinetic term is
\begin{equation}
{\cal L}^{(k)}_{GN} = \frac{f_1^{\mu\nu}}{2} (\partial_\mu\rho_j\partial_\nu \rho_j)
+ \frac{f^{\mu\nu}_2}{\rho^2}(\rho_j\partial_\mu\rho_j)(\rho_i
\partial_\nu\rho_i)
\label{eq9}
\end{equation}
where $\mbox{\boldmath$\rho$}=(\sigma,\pi)$ and $f^{\mu\nu}_1$, 
$f^{\mu\nu}_2$ are functions of $\rho^2$. To find the functions 
$f^{\mu\nu}_1$ and $f^{\mu\nu}_2$, one can use different methods. We
utilize the same method as in Ref.~\cite{4} (see Appendix A in that
paper). The result is:
\begin{eqnarray}
f^{\mu\nu}_1(\rho^2)&=&-\frac{i}{2}
\int\frac{d^2k}{(2\pi)^2} tr\left[
S(k)i\gamma_5\frac{\partial^2S(k)}{\partial k_{\mu}\partial k_{\nu}}
i\gamma_5\right],
\label{eq10} \\
f^{\mu\nu}_2(\rho^2)&=&-\frac{i}{4}
\int\frac{d^2k}{(2\pi)^2} tr\left[
S(k)\frac{\partial^2S(k)}{\partial k_{\mu}\partial k_{\nu}}-
S(k)i\gamma_5\frac{\partial^2S(k)}{\partial k_{\mu}\partial k_{\nu}}
i\gamma_5\right],
\label{eq11}
\end{eqnarray}
with  $S(k)=i(k^{\mu}\gamma_{\mu}+\rho)/(k^2-\rho^2)$. The
explicit form of these functions is:
\begin{equation}
f^{\mu\nu}_1=g^{\mu\nu}\frac{\tilde{N}_c}{4\pi\rho^2},\qquad
f^{\mu\nu}_2=-g^{\mu\nu}\frac{\tilde{N}_c}{12\pi\rho^2}
\label{eq12}
\end{equation}

\section{The Interplay between the GN Model and the 
NJL Model in a Magnetic Field}

In this section, we compare the effective actions in the GN model and
in the NJL model in a magnetic field, and we establish a rather
interesting connection between these two models.

The analog of the Lagrangian density (\ref{eq2}) in the NJL model in
a magnetic field is 
\begin{equation}
{\cal L}^{\prime}=\frac{1}{2} 
    \left[\bar{\Psi}, (i\gamma^\mu  D_\mu) \Psi\right] 
- \bar{\Psi}(\sigma+i\gamma^5\pi)\Psi
- \frac{1}{2G} \left(\sigma^2+\pi^2\right)
\label{eq2njl}
\end{equation}
where $D_\mu=\partial_\mu - ie A_\mu^{ext}$, 
$A_\mu^{ext}=Bx^2\delta_\mu^3$
(the magnetic field is in $+x_1$ direction).

In leading order in $1/N_c$, the effective action 
in the NJL model in a magnetic field is derived
in Refs.~\cite{2,4}. The effective potential and the kinetic term are
($\mbox{\boldmath$\rho$}=(\sigma,\pi)$):
\begin{eqnarray}
V(\rho)&=&\frac{\rho^2}{2G}+\frac{N_c}{8\pi^2} 
       \Bigg[\frac{\Lambda^4}{2}
     + \frac{1}{3l^4}\ln(\Lambda l)^2 
     + \frac{1-\gamma-\ln2}{3l^4}-(\rho\Lambda)^2
     + \frac{\rho^4}{2}\ln(\Lambda l)^2  
 \nonumber\\
 &+&\frac{\rho^4}{2}(1-\gamma-\ln2)
     + \frac{\rho^2}{l^2}\ln\frac{(\rho l)^2}{2}
     - \frac{4}{l^4}\zeta^{\prime} 
       (-1,\frac{(\rho l)^2}{2}+1)\Bigg]
    + O\left(\frac{1}{\Lambda}\right),
\label{eq13}
\end{eqnarray}
\begin{equation}
{\cal L}^{(k)} = \frac{f_1^{\mu\nu}}{2} 
         (\partial_\mu\rho_j\partial_\nu \rho_j)
       + \frac{f^{\mu\nu}_2}{\rho^2}
         (\rho_j\partial_\mu\rho_j)(\rho_i \partial_\nu\rho_i)
\label{eq14}
\end{equation}
with $f^{\mu\nu}_1$ and 
     $f^{\mu\nu}_2$ 
being diagonal tensors:
\begin{eqnarray}
f^{00}_1 = - f^{11}_1&=&\frac{N_c}{8\pi^2}
           \left[\ln\frac{(\Lambda l)^2}{2}
         - \psi\left(\frac{(\rho l)^2}{2}+1\right)
         + \frac{1}{(\rho l)^2} - \gamma
         + \frac{1}{3} \right], 
         \nonumber\\
f^{22}_1 = f^{33}_1&=& - \frac{N_c}{8\pi^2}
           \left[\ln\frac{\Lambda^2}{\rho^2}
         - \gamma+\frac{1}{3}\right], 
         \nonumber\\
f^{00}_2 = - f^{11}_2&=&-\frac{N_c}{24\pi^2}
           \left[\frac{(\rho l)^2}{2}
           \zeta\left(2,\frac{(\rho l)^2}{2}+1\right)
         + \frac{1}{(\rho l)^2}\right], 
         \label{eq15} \\
f_2^{22} = f^{33}_2&=& - \frac{N_c}{8\pi^2}
    \Bigg[(\rho l)^4 \psi \left(\frac{(\rho l)^2}{2}+1\right)
         - 2(\rho l)^2\ln\Gamma\left(\frac{(\rho l)^2}{2}\right)
         \nonumber\\
        &&\qquad - (\rho l)^2 \ln\frac{(\rho l)^2}{4\pi}
        - (\rho l)^4 - (\rho l)^2+1\Bigg].
\nonumber
\end{eqnarray}
Here $G$ is the NJL coupling constant, $N_c$  is the number of
colors, $\zeta(\nu,x)$ is the generalized Riemann zeta function,
$\zeta^{\prime}(\nu,x)=\partial\zeta(\nu,x)/\partial\nu$,  
$\gamma\approx0.577$ is the Euler constant,
$\psi(x)=d\left(\ln\Gamma(x)\right)/dx$,
and $l\equiv|eB|^{-1/2}$ is the magnetic length.
The gap equation 
$dV/d\rho=0$ is\footnote{We consider the case of a
large ultraviolet cutoff: $\Lambda^2\gg\bar{\sigma}^2$, $|eB|$, 
where $\bar{\sigma}$ is a minimum of the potential $V$.}
\begin{equation}
\rho\Lambda^2  \left(\frac{1}{g}-1\right)=
        - \rho^3\ln\frac{(\Lambda l)^2}{2}+\gamma\rho^3
        + \frac{\rho}{l^2}\ln\frac{(\rho l)^2}{4\pi}
        + \frac{2\rho}{l^2}\ln\Gamma\left(\frac{(\rho l)^2}{2}\right)
        + O\left(\frac{1}{\Lambda}\right),
\label{eq16}
\end{equation}
where the dimensionless coupling constant $g=N_cG\Lambda^2/4\pi^2$.
In the derivation of this equation, we used the relations \cite{10}:
\begin{eqnarray}
\frac{\partial}{\partial x}\zeta(\nu,x)=-\nu\zeta(\nu+1,x),&& 
                            \label{eq17} \\
\left.\frac{\partial}{\partial \nu}\zeta(\nu,x)\right|_{\nu=0}=
              \ln\Gamma(x)-\frac{1}{2}\ln2\pi,
           &\quad& \zeta(0,x)=\frac{1}{2}-x.
\label{eq18}
\end{eqnarray}
    
As $B\to0$ ($l\to\infty$), we recover the known gap equation in the NJL
model (for a review see Ref.\cite{11}):
\begin{equation}
\rho\Lambda^2\left(\frac{1}{g}-1\right)=
          - \rho^3\ln\frac{\Lambda^2}{\rho^2}.
\label{eq19}
\end{equation}
This equation admits a nontrivial solution only if $g$ is supercritical,
$g>g_c=1$ (as Eq.(\ref{eq2njl}) implies, a solution to the gap equation, 
$\rho=\bar{\sigma}$, coincides with the fermion dynamical mass, 
$\bar{\sigma}=m_{dyn}$). As was shown in Refs.~\cite{2,4}, 
at $B\neq0$, a non--trivial solution exists for all $g>0$.

Let us consider the case of small subcritical $g$, $g\ll g_c=1$, in
detail. A solution is seen to exist for this case if $\rho l$ is
small. Specifically, for $g\ll 1$, the left--hand side of 
Eq.(\ref{eq16}) is positive. Since the first term of the 
right--hand side in this equation is negative, we conclude 
that a non--trivial solution to this equation may exist only for
\begin{equation} 
\rho^2 \ln(\Lambda l)^2 \ll 
       \frac{1}{l^2} \ln\frac{1}{(\rho l)^2}
\label{eq20}
\end{equation}
($\Gamma(\rho^2 l^2/2)\approx 2/(\rho l)^2$). 
We then find the solution:
\begin{equation}
m^2_{dyn} \equiv \bar{\sigma}^2 = \frac{|eB|}{\pi}\exp
          \left(-\frac{4\pi^2(1-g)}{|eB| N_c G}\right) 
          = \frac{|eB|}{\pi}\exp
          \left(-\frac{(1-g)\Lambda^2}{g|eB|}\right).
\label{eq21}
\end{equation}
Actually, since Eq.(\ref{eq21}) implies that condition (\ref{eq20}) is
violated only if $(1-g) < |eB|/\Lambda^2$, the expression (\ref{eq21}) is
valid for all $g$ outside that (scaling) region near the critical
value $g_c=1$. Note that in the scaling region ($g\to g_c-0$) the
expression for $m_{dyn}^2$ is different \cite{2,4}:
\begin{equation}
m_{dyn}^2\simeq |eB|
            \frac{\ln\left[(\ln\Lambda^2l^2)/\pi \right]}
                 {\ln\Lambda^2l^2}.
\label{eq22}
\end{equation}
Let us compare relation (\ref{eq21}) with relation (\ref{eq8}) for the
dynamical mass in the GN model. The similarity between them is
evident: $|eB|$ and $|eB|G$ in Eq.(\ref{eq21}) play the role of an
ultraviolet cutoff and the dimensionless coupling constant $\tilde{G}$
in Eq.(\ref{eq8}). Let us discuss this connection and show that it is
intimately connected with the dimensional reduction $3+1\to 1+1$ in
the dynamics of the fermion pairing in a magnetic field.

Eq.(\ref{eq8}) implies that the GN model is asymptotically free, with
the bare coupling constant
$\tilde{G}=2\pi/\tilde{N}_c\ln(\Lambda^2/m_{dyn}^2)\to 0$ as
$\Lambda\to\infty$. Let us now consider the following limit in the
NJL model in a magnetic field: $|eB|\to\infty$, $|eB|/\Lambda^2=b\ll
1$. Then relation  (\ref{eq21}), which can be rewritten as
\begin{equation}
m^2_{dyn} = \frac{b\Lambda^2}{\pi}
            \exp\left(-\frac{(1-g)}{bg}\right),
\label{eq23}
\end{equation}
implies that the behavior of the bare coupling constant $g$ must be 
\begin{equation}
      g \simeq \frac{1}{b\ln(b\Lambda^2/\pi m_{dyn}^2)} \to 0,
\label{eq24}
\end{equation}
in order to get a finite value for $m_{dyn}^2$ in this
limit. Thus in this ``continuum" limit, we recover the same behavior for 
the coupling $g$ in the NJL model as for the coupling constant
$\tilde{G}$ in the GN model.

Let us now compare the effective potentials in these two models. At
first glance, the expressions (\ref{eq6}) and (\ref{eq13}) for the
effective potentials in these models look very
different: the character of ultraviolet divergences in
1+1 and 3+1 dimensional theories 
is essentially different. However,
using Eqs.(\ref{eq17}) and (\ref{eq18}), the expression (\ref{eq13}) can be
rewritten, for small $\rho l$, as 
\begin{eqnarray}
V(\rho) &=& V(0) + \frac{N_c\rho^2}{8\pi^2l^2} 
    \Bigg[ (\Lambda l)^2 \left(\frac{1}{g}-1\right)
        - 1 + \ln(\pi\rho^2 l^2) +
    \nonumber\\
        && \qquad + \frac{(\rho l)^2}{2}\ln\frac{(\Lambda l)^2}{2}
        + O((\rho l)^4)\Bigg].
\label{eq25}
\end{eqnarray}
Then, expressing the coupling constant $g$ through $m_{dyn}$ from
Eq.(\ref{eq21}), we find that
\begin{eqnarray}
V(\rho) &=& V(0) + \frac{N_c|eB|}{8\pi^2}\rho^2
        \left[ \ln\frac{\rho^2}{m_{dyn}^2} - 1 + O((\rho l)^2) \right] .
\label{eq26}
\end{eqnarray}
Here we used the fact that, because of Eq.(\ref{eq20}), the ratio 
$(\rho l)^2$ is small near the minimum $\rho=m_{dyn}$.

The expressions (\ref{eq7}) and (\ref{eq26}) for the potentials in these two
models look now quite similar. There is however an additional factor
$|eB|/2\pi$ in the expression (\ref{eq26}). Moreover, the  field 
$\rho$, which depends on the two coordinates $x_0$ and
$x_1$ in the GN model, depends on the four coordinates 
$x_0$, $x_1$, $x_2$ and $x_3$ in the NJL model.

In order to clarify this point, let us turn to the analysis of the
kinetic term (\ref{eq14}) in the effective action of the NJL model
in a magnetic field.

Because of the expression (\ref{eq21}) for $m_{dyn}$ at small $g$, the
term $1/(\rho l)^2$ dominates in the functions $f_{1}^{00}=-f_{1}^{11}$ 
and $f_{2}^{00}=-f_{2}^{11}$, around the minimum $\rho=m_{dyn}$:
\begin{eqnarray}
f^{00}_1=-f^{11}_1 \simeq
        \frac{N_c}{8\pi^2}\frac{1}{\rho^2l^2},\quad
f^{00}_2=-f^{11}_2 \simeq
        -\frac{N_c}{24\pi^2}\frac{1}{\rho^2l^2}.
\label{eq27}
\end{eqnarray}
Up to the additional factor $|eB|/2\pi$, these functions coincide
with those in (\ref{eq12}) in the GN model.
On the other hand, the functions $f^{22}_1=f^{33}_1$ and
$f^{22}_2=f^{33}_2$, connected with derivatives with respect to the
transverse coordinates, are strongly supressed, as compared to the
functions (\ref{eq27}), and the ratios of the functions
$f^{22}_1=f^{33}_1$ and $f^{22}_2=f^{33}_2$ to those in
(\ref{eq27}) go rapidly (as $m^{2}_{dyn}/\Lambda^2$) to zero as 
$|eB|\to \infty$, $|eB|/\Lambda^2=b$.                                 

As a result, the coordinates $x_2$ and $x_3$ become redundant
variables in this limit: there are no transitions of field quanta
between different points in the $x_{2}x_{3}$--plane. Therefore
the model degenerates into a set of independent $(1+1)$--dimensional
models, labeled by ${\bf x}_{\perp}=(x_2,x_3)$ coordinates.
Let us show that all these models coincide with a $(1+1)$--dimensional
GN model with the number of colors
$\tilde{N}_c=(b\pi/2)N_c$, $b=|eB|/\Lambda^2$. 

Let us put the NJL model 
on a lattice with $a$ the lattice 
spacing of the discretized space-time (in Euclidean region).
Then its effective action can be
written as
\begin{eqnarray}
\Gamma_{NJL}(\sigma, \pi) &=& \int dx_2 dx_3 \int dx_4 dx_1 L_{NJL}^{(eff)}
(\sigma(x), \pi(x)) \nonumber  \\ 
&\simeq& \frac{1}{2\pi} |eB| a^4 \sum 
\limits_{i,j = - \infty}^{\infty} \sum \limits_{n,m = - \infty}^{\infty}  
\tilde{L}_{NJL} ^{(eff)}(\sigma_{ij}(n,m), \pi_{ij}(n,m)) 
\label{eq28} 
\end{eqnarray}            
where $\sigma_{ij}(n,m) = \sigma(x)$, $\pi_{ij}(n,m) = \pi(x)$, with
$x_2=ia$, $x_3=ja$, $x_4=ix_0=na$, $x_1=ma$, and here the factor
$|eB|/2\pi$ was explicitly factorized from $L_{NJL}^{(eff)}$.
Now, taking into account Eqs.(\ref{eq7}),(\ref{eq12}) and Eqs.(\ref{eq27}),
(\ref{eq28}), we find that 
\begin{eqnarray}
\Gamma_{NJL} &=& \frac{1}{2\pi} |eB| a^4 \sum \limits_{i,j =
- \infty}^{\infty} \sum \limits_{n,m = - \infty}^{\infty}  \tilde{L}_{NJL}
^{(eff)}(\sigma_{ij}(n,m), \pi_{ij}(n,m)) \nonumber \\
&\to& \sum 
\limits_{x_2, x_3} \int dx_4 dx_1 L_{GN}^{(eff)} (\sigma_{x_2 x_3} 
(x_{\parallel}), \pi_{x_2 x_3} (x_{\parallel})) 
\label{eq29} 
\end{eqnarray} 
in the ``continuum" limit with
$(|eB|/2\pi)a^2\equiv\pi|eB|/2\Lambda^2=
b\pi/2$ (here $\Lambda=\pi/a$ is the ultraviolet cutoff on the
lattice)\footnote{Of course, the ultraviolet cutoff on the lattice is
different from the cutoff in the proper-time regularization used above.
However, since the constant $b=|eB|/\Lambda^2$ is anyway arbitrary here,
we use the same notation for the cutoff on the lattice as in the 
proper-time regularization.}.
The lagrangian density ${\cal L}^{(eff)}_{GN}$ corresponds to
the GN model with the number of colors $\tilde{N}_c=(\pi/2C)N_c$. Note
that the symbol $\sum \limits_{x_2, x_3}$
here is somewhat formal and it just implies that the GN
model occurs at each point in the $x_{2}x_{3}$--plane.

The physical meaning of this reduction of the
NJL model in a magnetic field is rather clear. At weak
coupling, the fermion pairing in a magnetic field takes place
essentially for fermions in the LLL with the momentum $k_1$=0.
The size of the radius of the LLL
orbit is $l$=$|eB|^{-1/2}$ \cite{12}. As the
magnetic field goes to infinity,
this
radius shrinks to zero. Then, because of the degeneracy in the LLL
\cite{12}, there are $(|eB|/2\pi)a^2N_c=(b\pi/2)N_c$ states
with $k_1$=0
at each point in the $x_{2}x_{3}$--plane. This degeneracy factor
is equal to $(|e\Phi|/2\pi)N_c$, where $\Phi$ is
the (local) magnetic flux across a plaquette on the lattice. It
leads to changing the number of colors, 
$N_c\to\tilde{N}_c=N_c(|e\Phi|/2\pi)$, in the GN model.
Note that since
$\tilde{N}_c$ appears analytically in the path integral of the theory,
one can give a non-perturbative meaning to the theory with
non-integral $\tilde{N}_c$.

A few comments are in order.

Since these GN models are independent, the parameters of the chiral
$U_{L}(1)\times U_{R}(1)$ transformations can depend on ${\bf x}_\perp$.
In other words, here the chiral group
is $\prod \limits_{x_{\perp}} U^{(x_\perp)}_{L}(1)\times
U^{(x_\perp)}_{R}(1)$. As a result, there are an infinite number
of gapless modes $\pi_{x_{2}x_{3}}(x_{\parallel})$ in the ``continuum"
limit.

Since there is no spontaneous breakdown of continuous symmetries
at $D=1+1$, the fields $\pi_{x_{2}x_{3}}(x_{\parallel})$ do not describe
NG bosons (though they do describe gapless BKT excitations) \cite{8}.

Since the magnetic field depends on $\Lambda$ in the ``continuum" 
limit, it can be considered as an additional parameter
(``coupling constant") in the renormalization group.
The ratio $b=|eB|/\Lambda^2$ is arbitrary here.
From the point of view of the renormalization group,
this can be interpreted as the presence of a line of 
ultraviolet fixed points for the dimensionless coupling $b$.
The values of $b$ on the line define the local magnetic flux
and, therefore, the number of colors $\tilde{N}_c$ in the
corresponding GN models. 

We emphasize that the reduction of the NJL model, described above,
takes place only as $|eB|\to\infty$. At finite
values of the magnetic field, the dynamics in the NJL and GN
models are different: while there is spontaneous chiral symmetry
breaking in the NJL model, the BKT phase is realized in the
GN model \cite{8}. The connection between these two sets of dynamics
is similar to that between the dynamics of 2--dimensional
and ($2+\epsilon$)--dimensional GN models.

Also, we emphasize that this discussion pertains only to the
NJL model with a weak coupling constant, when relation (\ref{eq21})
is valid. In the case of the NJL model with a near--critical $g$, the
situation is different: when $g\to g_c-0$, (\ref{eq22}) is valid. 
The difference between these two dynamical regimes reflects the
fact that, while at weak coupling the LLL dominates, at
near--critical $g$ all Landau levels are relevant \cite{4}.

\section{Magnetic Catalysis and the Aharonov-Bohm Effect.}

The relation between $\tilde{N}_c$ and $N_c$ obtained
in the previous section is:
\begin{equation}
\tilde{N}_c=N_c (|e\Phi|/2\pi).
\label{eqN_c1}
\end{equation}
If we asserted that $\tilde{N}_c$ is a (non-negative) integer,
we would be led to the
quantization condition for $N_c(\Phi/2\pi)$ coinciding with
the quantization condition for a magnetic flux,
$q(\Phi/2\pi)=n$, 
at which the Aharonov-Bohm (AB) scattering 
of particles with the electric charge
$q=eN_c$ disappears \cite{AB}.

Let us discuss this point in more details.

The AB effect is characterized by the following features:
\begin{enumerate}
\item When $q(\Phi/2\pi)$ is not integer, there is a nontrivial 
scattering of particles with the charge $q$ in a
line-like (or point-like, in 2+1 dimensions) solenoid field.
The effect looks as a non-local one: particles being
outside of the solenoid somehow feel the magnetic field 
inside it.

The effect is intimately connected with the boundary
conditions for the wave functions
of particles on the solenoid surface.
\item When $q(\Phi/2\pi)$ is integer, the magnetic flux in
a line-like solenoid is not observable.
\end{enumerate}

Let us now turn to the dimensional reduction in the NJL
model in a magnetic field. The number of colors $\tilde{N_c}$ in
the corresponding GN model is given by expression (\ref{eqN_c1}).
Its physical meaning is clear: it is just the number of degrees
of freedom connected with one plaquette on the lattice described 
in the previous section. Now, has this number to be integer?
The answer is ``no'': the plaquette is just a part
of the lattice, and there is no such an additional constraint
in the system.

When this number is non-integer,
the dynamics of the plaquette is not completely
factorized from the dynamics of the rest of the lattice
even in the ``continuum'' limit considered in
the previous section. In
other words, those degrees of freedom are not confined inside
one plaquette (that has to be reflected in a nontrivial
boundary condition  for the fields on the plaquette
boundary). As a result,
even in the ``continuum'' limit 
the dynamics of a plaquette is non-local in this case: indeed, it
is described by a (1+1) dimensional GN model with a
non-integer $\tilde{N_c}$, which certainly is not a local
field theory.

These features are similar to the features of the AB effect 
discussed in item 1 above.

On the other hand, for integer $\tilde{N_c}$ the dynamics
connected with one plaquette does factorize from the dynamics 
of the rest of the lattice in the ``continuum'' limit, and 
therefore it 
is described by a local GN model. Like in the case of the
magnetic flux satisfying the AB condition (see item 2),
the flux connected
with a plaquette is not observable (for a (1+1)-dimensional
observer) in this case.

Therefore we are led to the
interpretation of the dynamics of the GN model
with a non-integer $\tilde{N_c}$ as a quantum field
theoretical analogue of the AB dynamics with a non-integer
$q(\Phi/2\pi)$.
The AB dynamics
occurs from the dynamics in a solenoid field
when the cross-section of the solenoid goes to zero.
Similarly, the (1+1)-dimensional GN model, with both
integer and non-integer $\tilde{N_c}$, occurs from the
(3+1)-dimensional NJL model in a magnetic field in the
``continuum'' limit, when the radius of the LLL shrinks to
zero.

It would be interesting to study Green's functions
in the GN model with a non-integer $\tilde{N_c}$.

\section*{Acknowledgments}


I thank V. P. Gusynin for useful remarks.
This work is based on the talk given at the Workshop
``Nonperturbative Methods in Quantum Field Theory''
(February 1998, Adelaide).
I am grateful to the members of the CSSN/NITP hosted by
the University of Adelaide, especially A. W. Thomas and
A. G. Williams, for their hospitality. The paper has been
finished while I was visiting Nagoya University.
My gratitudes to
K. Yamawaki for his hospitality. I wish to acknowledge
the Monbusho (Ministry of Education of Japan) for its
support during my stay in Nagoya University.


\begin{thebibliography}{99}

\bibitem[*]{email}
            E-mail: {\tt miransky@gluk.apc.org}

\bibitem{1} V.P.Gusynin, V.A.Miransky, and I.A.Shovkovy,
{\sl Phys. Rev. Lett.} {\bf 73,} 3499 (1994);
{\sl Phys. Rev.} D{\bf 52,} 4718 (1995).

\bibitem{2} V.P.Gusynin, V.A.Miransky, and I.A.Shovkovy,
{\sl Phys. Lett.} B{\bf 349,} 477 (1995).

\bibitem{3} V.P.Gusynin, V.A.Miransky, and I.A.Shovkovy,
{\sl Phys. Rev.} D{\bf 52,} {4747} (1995).

\bibitem{4} V.P.Gusynin, V.A.Miransky, and I.A.Shovkovy,
{\sl Nucl. Phys.} B{\bf 462,} 249 (1996).

\bibitem{ng} C. N. Leung, Y. J. Ng, and A. W. Ackley,
{\sl Phys. Rev.} D{\bf 54,} {4181} (1996);
D. K. Hong, Y. Kim, and S. Sin,
{\sl Phys. Rev.} D{\bf54,} {7879} (1996);
I. A. Shovkovy and V. M. Turkovsky,
{\sl Phys. Lett.} B{\bf 367,} {213} (1996);
D. M. Gitman, S. D. Odintsov, and Yu. I.
Shil'nov, {\sl Phys. Rev.} D{\bf 54,} {2968} (1996);
T. Inagaki, T. Muta, and S. D. Odintsov, {\sl Prog.
Theor. Phys. Suppl}. {\bf 127,} 93 (1997);
E. Elizalde, Yu. I. Shil'nov, and V. V. Chitov,
{\sl Class. Quant. Grav.} {\bf 15,} 735 (1998);
A. V. Shpagin, {\em Dynamical mass
generation in (2+1)-dimensional electrodynamics in
external magnetic field}, hep-ph/9611412;
I. A. Shushpanov and A. V. Smilga,
{\sl Phys. Lett.} B{\bf 402,} {351} (1997);
D. Ebert and V. Ch. Zhukovsky,
{\sl Mod. Phys. Lett.} A{\bf 12,} 2567 (1997);
D. K. Hong, {\sl Phys. Rev.} D{\bf 57,} {3759} (1998);
V. P. Gusynin, D. K. Hong, and
I. A. Shovkovy, {\sl Phys. Rev.} D{\bf 57,} {5230} (1998);
A. Yu. Babansky, E. V. Gorbar, and
G. V. Shchepanyuk, {\sl Phys. Lett.} B{\bf 419,} {272} (1998);
S. Kanemura, H.-T. Sato, and H. Tochimura, {\sl Nucl. Phys.}
B{\bf 517,} {567} (1998);
K. Farakos, G. Koutsoumbas, and N. E.
Mavromatos, {\em Dynamic flavour symmetry breaking by
a magnetic field in lattice QED3}, hep-lat/9802037;
E. J. Ferrer and V. de la Incera, {\em 
Ward-Takahashi identity with external field in ladder
QED}, hep-th/9803226.

\bibitem{el} V. Elias, D. G. C. McKeon, V. A. Miransky,
and I. A. Shovkovy, {\sl Phys. Rev.} D{\bf 54,} {7884} (1996).

\bibitem{ngT} D.-S. Lee, C. N. Leung, and Y. J. Ng,
{\sl Phys. Rev.} D{\bf 55,} {6504} (1997); {\sl Phys. Rev.}
D{\bf 57,} {5224} (1998).

\bibitem{gs} V. P. Gusynin and I. A. Shovkovy,
{\sl Phys. Rev.} D{\bf 56,} {5251} (1997).

\bibitem{klev} S. Kawati, G. Konisi, and H. Miyata,
{\sl Phys. Rev.} D{\bf28,} {1537} (1983); S. P. Klevansky and
R. H. Lemmer, {\sl Phys. Rev.} D{\bf 39,} {3478} (1989).

\bibitem{krive} I. V. Krive and S. A. Naftulin,
{\sl Phys. Rev.} D{\bf 46,} {2737} (1992); K. G. Klimenko, 
{\sl Z. Phys.} C{\bf 54,} {323} (1992).

\bibitem{nick} K. Farakos and N. E. Mavromatos, {\em
Gauge theory approach to planar doped antiferromagnets
and external magnetic fields}, cond-matt/9710188;
N. E. Mavromatos and A. Momen, {\em Induced magnetic
moments in three dimensional gauge theories with
external magnetic fields}, hep-th/9802119.

\bibitem{ssw} G. W. Semenoff, I. A. Shovkovy, and
L. C. R. Wijewardhana, {\em Phase transition induced by
a magnetic field}, hep-ph/9803371.

\bibitem{inc1} E. J. Ferrer and V. de la Incera,
{\em Yukawa interactions and dynamical generation of
mass in an external magnetic field}, hep-th/9805103.

\bibitem{itoh} T. Itoh and H. Kato, {\em Dynamical
generation of fermion mass and magnetic field in three
dimensional QED with Chern-Simons term}, hep-th/9802101
(to appear in {\sl Phys. Rev. Lett}).

\bibitem{hosot} Y. Hosotani, {\sl Phys. Lett.}
 B{\bf 319,} {332} (1993);
{\sl Phys. Rev.} D{\bf 51,} {2022} (1995). 

\bibitem{5} D. Gross and A. Neveu, {\sl Phys. Rev.} D{\bf 10,} 3235 
(1974).

\bibitem{AB} Y. Aharonov and D. Bohm, {\sl Phys. Rev.}
{\bf 115,} 485 (1959).

\bibitem{7} N.D. Mermin and H. Wagner, 
{\sl Phys. Rev. Lett.} {\bf 17,}
1133 (1966); S. Coleman, 
{\sl Commun. Math. Phys.} {\bf 31,} 259 (1973).

\bibitem{8} E. Witten, {\sl Nucl. Phys.} B{\bf 145,} 110 (1978).

\bibitem{10} I.S. Gradshtein and I.M. Ryzhik, {\em Table of 
Integrals, Series and Products} (Academic Press, Orlando, 1980).

\bibitem{11} V.A. Miransky, {\sl Dynamical Symmetry Breaking in 
Quantum Field Theories} (World Scientific, Singapore, 1993).

\bibitem{12} A.I. Akhiezer and V.B. Berestetsky, {\em Quantum
Electrodynamics} (Interscience, NY, 1965).


\end{thebibliography}
\end{document}